\documentclass[aps,prd,amssymb,showpacs,floatfix,preprint,nofootinbib]{revtex4-1}
\usepackage{amsmath,amsfonts,graphics,epsfig,amssymb,color}

\begin{document}
\title{Schwarzian functional integrals calculus }

\author{Vladimir V. Belokurov}

\email{vvbelokurov@yandex.ru}

\affiliation{Lomonosov Moscow State University, Institute for Nuclear
Research of  Russian Academy of Sciences
}

\author{Evgeniy T. Shavgulidze}

\email{shavgulidze@bk.ru}

\affiliation{Lomonosov Moscow State University}

%\pacs{}

%\preprint{}
%\date{}

\begin{center}
\begin{abstract}
We derive the general rules of functional integration in the theories of Schwarzian type,
thus completing the elaboration of Schwarzian functional integrals calculus initiated in \cite{(BShExact)}, \cite{(BShCorrel)}.

Our approach is  mathematically rigorous and does not contain any unproved conjectures.
It is based on the analysis of the properties of the measures on the groups of diffeomorphisms, and does not appeal for the experience from other physical models.

Its great merit consists in reducing a problem of functional integration to that of the only functional integral (\ref{E})
that is calculated explicitly with the result written in the form of the ordinary integral.

We evaluate two-point and four-point correlation functions defined as functional integrals over the groups $Diff^{1}_{+}(\textbf{R}) $ and $Diff^{1}_{+}(S^{1})\,,$ and discuss the difference between the results in the two cases.

\end{abstract}
\end{center}
\maketitle

\vspace{1cm}
\section{ Introduction}
\label{sec:intr}

In recent years,
it has become clear that a quantum mechanical model of Majorana fermions with a random interaction (Sachdev-Ye-Kitaev model), the holographic
description of the Jackiw-Teitelboim dilaton gravity, open string theory and some other models lead to the same effective  theory with the Schwarzian action
\begin{equation}
   \label{Act1}
  I_{Schw}= -\frac{1}{\sigma^{2}}\int \limits _{S^{1}}\, \mathcal{S}ch  \{f,\,t\}\,dt\,,
\end{equation}
where $S^{1} $ is the unit circle, and
$
\mathcal{S}ch  \{f,\,t\}=
\left(\frac{f''(t)}{f'(t)}\right)'
-\frac{1}{2}\left(\frac{f''(t)}{f'(t)}\right)^2
$ is the Schwarzian derivative.

In some approximations, these physical models turn out to be reparametrization-invariant. And it is the reason of the appearance of the common effective dynamics given by the $SL(2,\textbf{R})$-invariant action (\ref{Act1}).

One can observe a real explosion of interest in the Schwarzian theory and the models leading to it, especially because of their expected chaotic behaviour.
The list of the papers studying the mentioned theories includes, but is not confined to, \cite{(SY)} - \cite{(GKRS)}, and it will not rest here.

Functional integrals in the Schwarzian theory are integrals of the form
\begin{equation}
   \label{GenForm}
 \int \limits_{\mathcal{F}}\,F\left(f,\,f'\right)\,
 \exp\left\{\frac{1}{\sigma^{2}}\int \limits _{S^{1}}\, \mathcal{S}ch\{f,\,t\}\,dt\, \right\}\,df\,,
\end{equation}
over the space $\mathcal{F}$ of one-time differentiable functions $f$ on the unit circle $(f'(t)>0)\,.$ After the substitution $\ f(t)= - \cot \pi\varphi (t)\,,$
it is turned into the group of diffeomorphisms\footnote{$\varphi$ is one-time differentiable and $\varphi'(t)>0\,.$}
$ Diff^{1}_{+}(S^{1})\,,$
and the integral (\ref{GenForm}) is written as
\begin{equation}
   \label{GenFormI}
\int \limits_{Diff^{1}_{+}(S^{1})}\,\tilde{\Psi}\left(\varphi,\,\varphi'\right)\,
  \exp\left\{\frac{1}{\sigma^{2}}\int \limits _{S^{1}}\, \left(\mathcal{S}ch\{\varphi,\,t\}+2\pi^{2}\,\left(\varphi'(t)\right)^{2}\right)\,dt\, \right\}\,d\varphi\,.
\end{equation}

In the SYK model and 2D gravity, correlation functions are written as functional integrals (\ref{GenForm}) or (\ref{GenFormI})
with $SL(2,\textbf{R})$-invariant integrands of the form
$\
\mathcal{O}\,\left(t_{1}\,,\,t_{2}\right)\cdots \mathcal{O}\,\left(t_{n-1}\,,\,t_{n}\right)\,,
$
where
\begin{equation}
   \label{O}
  \mathcal{O}_{f}\left(t_{1}\,,\,t_{2}\right)=\frac{\left(f'(t_{1})f'(t_{2})\right)^{\frac{1}{4}}}{\left |f( t_{2})-f(t_{1})\right|^{\frac{1}{2}}}\,,\ \ \ \
or \ \ \ \
\mathcal{O}_{\varphi}\left(t_{1}\,,\,t_{2}\right)=\frac{\left(\varphi'(t_{1})\varphi'(t_{2})\right)^{\frac{1}{4}}}{\left |\sin \left[\pi \varphi ( t_{2})-\pi\varphi(t_{1})\right]\right|^{\frac{1}{2}}}\,.
\end{equation}

The group nature of the space of integration may lead someone into temptation to consider some mathematical constructions as invariant measures.
However, it is well known that
invariant measures analogous to the Haar measure on finite-dimensional groups do  not exist for noncompact groups \cite{(Weil)}.

In particular, the
"continual product" (see, e.g., \cite{(Polyakov)}, \cite{(ASh)})
$$
\prod\limits_{x} \frac{df(x)}{f'(x)}
$$
is a symbolic form of a generalized left-invariant Haar measure on the group of diffeomorphisms.
It is not a countably additive measure in the rigorous sense! And all the results obtained by manipulating this expression have a heuristic character only.
Here, one has a complete analogy with nonexistence of the Lebesgue measure on infinite-dimensional linear spaces. In this case, one is to use Gaussian measures (for example, the Wiener measure) \cite{(Kuo)}.

The authors of the seminal papers \cite{(BAK1)}, \cite{(MTV)}, \cite{(SW)}, as well as of also remarkable the more recent ones \cite{(KitSuh2)}, \cite{(Yang)}, reduced the problem, in some way or another, to integration over the Wiener measure.

In particular, in \cite{(BAK1)}, \cite{(BAK2)}, Schwarzian  functional integrals were represented as sums over the eigenstates of the 1D Liouville model \cite{(Tesch)}. In \cite{(MTV)}, \cite{(KitSuh2)}, \cite{(Yang)}, and some other papers, the eigenstates of the Hamiltonian of a particle on the hyperbolic upper-half plane in a constant magnetic (electric) field \cite{(Comt1)}, \cite{(Comt2)} were used for this purpose.

Another promising approach is to re-formulate the problem in terms of a random matrix theory \cite{(Cotler)}, \cite{(SSS2)}, \cite{(SW2)}.

It is worth mentioning also the papers
 \cite{(BShUnusual)}, \cite{(BShPolar)} where we expressed functional integrals in the Schwarzian theory as corresponding
 path integrals in conformal quantum mechanics and
 evaluated some nontrivial functional integrals in both theories.

However the goal of this paper is to develop  the approach to  direct functional integration based on the quasi-invariance of the measure on the group of diffeomorphisms.

Although there is no Haar measure on the group of diffeomorphisms $ Diff^{1}_{+}(S^{1})$,  there exists countably additive measure quasi-invariant under the action of the subgroup $Diff^{3}_{+}(S^{1})$ \cite{(Shavgulidze1978)}-\cite{(Shavgulidze2000)}\footnote{
The quasi-invariance means that under the action of the subgroup $Diff^{3}_{+}(S^{1})$,  the measure transforms to itself multiplied by a function parametrized by the elements of the subgroup (the Radon-Nikodim derivative).}.

Formally the measure on $ Diff^{1}_{+}(S^{1})$ can be written as\footnote{
Note that the second and third derivatives in (\ref{MeasureS}) are understood in a generalized sense, similar to the first derivative
in the formal expression for the Wiener measure on the space of continuous functions
\begin{equation}
   \label{WienerS}
 w_{\sigma}(d\xi)=\exp\left\{ -\frac{1}{2\sigma^{2}}\int \limits _{S^{1}}\,\left(\xi'(t) \right)^{2}dt \right\}\ d\xi\,.
\end{equation} }
\begin{equation}
   \label{MeasureS}
   \mu_{\sigma}(d\varphi)=\exp\left\{\frac{1}{\sigma^{2}}\int \limits _{S^{1}}\, \mathcal{S}ch\{\varphi,\,t\}\,dt  \right\}  d\varphi\,.
\end{equation}

The measure $\mu_{\sigma}(d\varphi) $ is generated by the Wiener measure under some special substitution of variables $\varphi=\varphi(\xi)$ (see the details in section \ref{sec:red}) .

In \cite{(BShExact)}, \cite{(BShCorrel)}, we evaluated some nontrivial functional integrals using the quasi-invariance of the measure (\ref{MeasureS}).
Here, we complete the elaboration of Schwarzian functional integrals calculus initiated in  \cite{(BShExact)}, \cite{(BShCorrel)} and illustrate it
evaluating functional integrals over the groups of diffeomorphisms  of the real axis $\left(Diff^{1}_{+}(\textbf{R})\right)\,, $ and of the circle $\left( Diff^{1}_{+}(S^{1})\right)$ that assign  two different types of two-point and four-point correlation functions correspondingly.

Our approach  is mathematically rigorous and does not contain any unproved conjectures.
On the other hand, it results in the very simple universal rules for a wide class of functional integrals over the groups of diffeomorphisms.

In section
\ref{sec:red}, we discuss the relation between the measure on the group $Diff^{1}_{+}([0,\,1])$ and the Wiener measure on the space of continuous functions $C([0,\, 1])\,,$ and explain how to reduce functional integrals over the group $ Diff^{1}_{+}(S^{1})$ to those over the group $Diff^{1}_{+}([0,\,1])\,.$
Then we obtain the rules of transformation of functional integrals over the group $Diff^{1}_{+}([0,\,1])$ with integrands of some special form into ordinary multiple integrals.

In section
\ref{sec:diffR},
the method developed in the previous section is applied to correlation functions defined as functional integrals
over the space $Diff^{1}_{+}(\textbf{R})\,. $

We demonstrate that those are precisely the integrals evaluated in \cite{(BAK1)}. Using our method we
 reproduce the results for correlation functions obtained in \cite{(BAK1)}, \cite{(MTV)}, and \cite{(KitSuh2)}, although written in a slightly different form.

In section \ref{sec:quasi-inv}, we demonstrate how the quasi-invariance of the measure  and the explicit form of the Radon-Nikodim derivative can be used to evaluate functional integrals.

In section \ref{sec:ren}, we explain how we exclude the infinite input of the noncompact group $SL(2,\textbf{R}) $ in Schwarzian functional integrals with $SL(2,\textbf{R})-$invariant integrands, and define functional integrals over the quotient space $Diff^{1}_{+}(S^{1})/SL(2,\textbf{R})\,. $

In section \ref{sec:correl},
we explicitly evaluate the functional integrals over $Diff^{1}_{+}(S^{1})/SL(2,\textbf{R}) $ for two-point and four-point correlation functions . The result is expressed in the form of ordinary multiple integrals that can be analyzed numerically.

  We demonstrate that all points of the circle $S^{1}$ give a nonzero input into the integrals for correlation functions. It leads to some nontrivial physical results. In particular, neither time-ordered nor out-of-time-ordered four-point correlation function is represented in the form of a product of two two-point correlation functions.

In section \ref{sec:concl}, we discuss the difference between correlation functions
obtained by functional integration over the group $ Diff^{1}_{+}(S^{1})$ and that over the group $Diff^{1}_{+}(\textbf{R})\,,$
and remark on its physical implications. Also we compare our results with those obtain in
 \cite{(BAK1)},  \cite{(MTV)}, \cite{(KitSuh2)}, \cite{(Yang)}.

Since   mathematical facts used here may be unfamiliar to some readers, to make our results easily verifiable we present the main steps of calculations in detail.
In this paper, we expand on our unpublished work \cite{(BShSimple)}.

\section{ Reduction of functional integrals over the group $Diff^{1}_{+}([0,1])$ to ordinary integrals}
\label{sec:red}

The rules of functional integration over the group $Diff^{1}_{+}(S^{1})$ are based on the study of functional integrals over the group of diffeomorphisms of the interval $Diff^{1}_{+}([0,1])\,.$
The measure on the group $Diff^{1}_{+}([0,1])$ as compared to (\ref{MeasureS}) has the additional factor depending on the ends of the interval:
\begin{equation}
   \label{Measure}
   \mu_{\sigma}(d\varphi)=\frac{1}{\sqrt{\varphi'(0)\varphi'(1)}} \exp\left\{ \frac{1}{\sigma^{2}}\left[   \frac{\varphi''(0)}{\varphi'(0)}-  \frac{\varphi''(1)}{\varphi'(1)}\right]\right\}\exp\left\{\frac{1}{\sigma^{2}}\int \limits _{0}^{1}\, \mathcal{S}ch \{\varphi,\,t\}\,dt  \right\}  d\varphi\,.
\end{equation}

The measure (\ref{Measure}) turns into the Wiener measure  $w_{\sigma}(d\xi)$ on $C_{0}([0,\, 1])$
\begin{equation}
   \label{Wiener}
 w_{\sigma}(d\xi)=\exp\left\{ -\frac{1}{2\sigma^{2}}\int \limits _{0}^{1}\,\left(\xi'(t) \right)^{2}dt \right\}\ d\xi\,.
\end{equation}
 under the following substitution of variables
\begin{equation}
   \label{phiksisubst}
 \varphi(t)=\frac{\int \limits _{0}^{t}\,e^{\xi(\tau)}d\tau}{\int \limits _{0}^{1}\,e^{\xi(\eta)}d\eta }  \,,\ \ \ \ \ \xi(t)=\log\varphi'(t)-\log\varphi'(0)\,,
\end{equation}
where $ \xi(t)$ is a continuous function on the interval $[0,\,1]\,$ satisfying the boundary condition
$\xi(0)=0\ \left(\xi\in C_{0}([0,\,1])\, \right)\,.$

Thus we obtain the relation between functional integrals (see also  \cite{(BShCorrel)})
\begin{equation}
   \label{IntEq}
  \int\limits_{Diff^{1}_{+} ([0,\,1]) }F\left(\varphi,\,\varphi'\right)\,\mu_{\sigma}(d\varphi)
 =\int\limits _{C_{0}([0,1]) }F\left(\varphi(\xi),\,(\varphi(\xi))'\right)\, w_{\sigma}(d\xi)\,.
\end{equation}

To evaluate functional integrals over the group $ Diff^{1}_{+}(S^{1})\,,$ it is convenient to represent them as integrals over the group $Diff^{1}_{+}([0,1])\,.$
   First we map the circle $S^{1}$ onto the interval $[0,\,1]$, and glue the ends of the interval by the equation
$\varphi'(0)=\varphi'(1)$ (or, using a figure of speech, construct an
\emph{ouroboros}\footnote{A snake biting its own tail.}).

In this case, the corresponding Wiener variable $\zeta  $ is a Brownian bridge $ \zeta(0)=\zeta(1)=0\,,\ \left(\zeta\in C_{00}([0,\,1])\, \right)\,,$ in contrast to the ordinary Wiener process $\xi$ in (\ref{IntEq}) with $\xi(0)=0$ and an arbitrary value of $\xi(1)\,.$ The conditional Wiener (Brownian) measure is related to the ordinary one by
\begin{equation}
   \label{Brown}
w_{\sigma}(d\xi)=w^{Brown}_{\sigma}(d\zeta)\frac{1}{\sqrt{2\pi}\sigma}\exp\left(-\frac{x^{2}}{2\sigma^{2}} \right)dx\,.
\end{equation}

Taking (\ref{IntEq}) and (\ref{Brown}) into account, we have
 the following equation:
\begin{equation}
   \label{Equality}
\frac{1}{\sqrt{2\pi}\sigma}\int\limits_{Diff^{1}_{+} (S^{1}) }F(\varphi,\,\varphi')\mu_{\sigma}(d\varphi)
=\int\limits_{Diff^{1}_{+} ([0,1]) }\delta\left(\frac{\varphi'(1)}{\varphi'(0)}-1 \right)\,F(\varphi,\,\varphi')\,\mu_{\sigma}(d\varphi)
\,.
\end{equation}

Now we consider the functional integral of the form
\begin{equation}
   \label{Int1}
\int\limits_{Diff^{1}_{+} ([0,\,1]) }\,\Phi(\varphi(t_{1});\,\varphi'(0),\,\varphi'(t_{1}),\,\varphi'(1))\,\mu_{\sigma}(d\varphi)\,.
\end{equation}
After the substitution (\ref{phiksisubst}), it is transformed into
\begin{equation}
   \label{IntEqFi1}
=\int\limits _{C_{0}([0,1]) }\,\Phi(\varphi(\xi(t_{1}));\,\varphi'(0),\,\varphi'(\xi(t_{1})),\,\varphi'(\xi(1)))\, w_{\sigma}(d\xi)\,.
\end{equation}

It is convenient to split the interval $[0,\,1]$ into the two intervals $[0,\,t_{1}]$ and $[t_{1},\,1]\,,$ and substitute
\begin{equation}
   \label{xieta}
\xi(t)=\eta_{0}\left(\frac{t}{t_{1}} \right)\,,\ \ \ 0\leq t\leq t_{1}\,; \ \ \ \ \ \ \ \ \ \ \ \xi(t)=\eta_{0}(1) +\eta_{1}\left(\frac{t-t_{1}}{1-t_{1}} \right)\,,\ \ \ t_{1}\leq t \leq 1\,.
\end{equation}
Now the exponent in the Wiener measure is written as
$$
\frac{1}{\sigma^{2}}\left(\int \limits _{0}^{t_{1}}\left(\xi'(t) \right)^{2}\,dt +\int \limits _{t_{1}}^{1}\left(\xi'(t) \right)^{2}\,dt\right)=\frac{1}{\sigma^{2}t_{1}}\int \limits _{0}^{1}\left(\eta'_{0}(t) \right)^{2}\,dt+\frac{1}{\sigma^{2}(1-t_{1})}\int \limits _{0}^{1}\left(\eta'_{1}(t) \right)^{2}\,dt\,.
$$
Due to the Markov property of  the Wiener process $\xi(t) \,,$ the measure $w_{\sigma}(d\xi)$ turns into the product of the two measures
$$
w_{\sigma \sqrt{t_{1}}}(d\eta_{0})\,w_{\sigma \sqrt{1-t_{1}}}(d\eta_{1})\,.
$$

To return to the integrals over the group of diffeomorphisms, define the functions $ \psi_{0}\,,\ \psi_{1}\ \in Diff^{1}_{+}([0,\,1])$
\begin{equation}
   \label{psi}
 \psi_{0}(t)=\frac{\int \limits _{0}^{t}e^{\eta_{0}(\tau)}d\tau }{\int \limits _{0}^{1}e^{\eta_{0}(\tau)}d\tau }\,, \ \ \ \ \ \ \ \ \ \ \ \ \psi_{1}(t)=\frac{\int \limits _{0}^{t}e^{\eta_{1}(\tau)}d\tau }{\int \limits _{0}^{1}e^{\eta_{1}(\tau)}d\tau }\,.
\end{equation}

In this way, we get
$$
\int\limits_{Diff^{1}_{+} ([0,\,1]) }\,\Phi(\varphi(t_{1});\,\varphi'(0),\,\varphi'(t_{1}),\,\varphi'(1))\,\mu_{\sigma}(d\varphi)
$$
\begin{equation}
   \label{IntEqFi2}
=\int\limits_{Diff^{1}_{+} ([0,\,1]) }\,
\int\limits_{Diff^{1}_{+} ([0,\,1]) }\,\Phi(\varphi(t_{1});\,\varphi'(0),\,\varphi'(t_{1}),\,\varphi'(1))\,\mu_{\sigma\sqrt{t_{1}}}(d\psi_{0})\,\mu_{\sigma\sqrt{1-t_{1}}}(d\psi_{1})
\,.
\end{equation}

The functions $\varphi$ and $\psi$ (and their derivatives) are related by the following equations:
\begin{equation}
   \label{FiPsi0}
\varphi(t)=\frac{t_{1}\,\psi'_{1}(0)\,\psi_{0}\left(\frac{t}{t_{1}} \right)}{t_{1}\psi'_{1}(0)+(1-t_{1})\psi'_{0}(1)}\,,\ \ \ \ \ 0\leq t\leq t_{1}\,;
\end{equation}
\begin{equation}
   \label{FiPsi1}
\varphi(t)=\frac{t_{1}\,\psi'_{1}(0)+(1-t_{1})\,\psi'_{0}(1)\,\psi_{1}\left(\frac{t-t_{1}}{1-t_{1}} \right)}{t_{1}\psi'_{1}(0)+(1-t_{1})\psi'_{0}(1)}\,,\ \ \ \ \ t_{1}\leq t \leq 1\,,
\end{equation}
and
\begin{equation}
   \label{DerFiPsi0}
\varphi'(t)=\frac{\,\psi'_{1}(0)\,\psi'_{0}\left(\frac{t}{t_{1}} \right)}{t_{1}\psi'_{1}(0)+(1-t_{1})\psi'_{0}(1)}\,,\ \ \ \ \ 0\leq t\leq t_{1}\,;
\end{equation}
\begin{equation}
   \label{DerFiPsi1}
\varphi'(t)=\frac{\psi'_{0}(1)\,\psi'_{1}\left(\frac{t-t_{1}}{1-t_{1}} \right)}{t_{1}\psi'_{1}(0)+(1-t_{1})\psi'_{0}(1)}\,,\ \ \ \ \ t_{1}\leq t \leq 1\,.
\end{equation}

In particular, the arguments in the integrand are
\begin{equation}
   \label{FiT1}
\varphi(t_{1})=\frac{t_{1}\,\psi'_{1}(0)}{t_{1}\psi'_{1}(0)+(1-t_{1})\psi'_{0}(1)}\,;\ \ \ \ \ \ \
\varphi' (t_{1})=\frac{\psi'_{0}(1)\psi'_{1}(0)}{t_{1}\psi'_{1}(0)+(1-t_{1})\psi'_{0}(1)}\,,
\end{equation}
\begin{equation}
   \label{DerFi01}
\varphi' (0)=\frac{\psi'_{0}(0)\psi'_{1}(0)}{t_{1}\psi'_{1}(0)+(1-t_{1})\psi'_{0}(1)}\,,
\ \ \ \ \ \ \
\varphi' (1)=\frac{\psi'_{0}(1)\psi'_{1}(1)}{t_{1}\psi'_{1}(0)+(1-t_{1})\psi'_{0}(1)}\,.
\end{equation}

Now  we can represent all functional integrals with integrands depending on $\varphi(t_{1}),\ \varphi'(0),$ $ \varphi'(t_{1}),\ \varphi'(1)$  in a similar way. To this end, we define the basic functional integral:
\begin{equation}
   \label{E}
\mathcal {E}_{\sigma}(u,\,v)=\int\limits_{Diff^{1} ([0,1]) }\,\delta\left(\varphi'(0)-u \right)\,\delta\left(\varphi'(1)-v \right)\,\mu_{\sigma}(d\varphi)\,,
\end{equation}
and rewrite the functional integral (\ref{IntEqFi2}) as
$$
\int\limits_{Diff^{1}_{+} ([0,\,1]) }\,
\int\limits_{Diff^{1}_{+} ([0,\,1]) }\,\Phi\left(\varphi(t_{1});\,\varphi'(0),\,\varphi'(t_{1}),\,\varphi'(1)\right)\,\mu_{\sigma\sqrt{t_{1}}}(d\psi_{0})\,\mu_{\sigma\sqrt{1-t_{1}}}(d\psi_{1})
$$
\begin{equation}
   \label{IntUV01}
=\int\limits_{0}^{+\infty}\cdot\cdot\cdot
\int\limits_{0}^{+\infty}du_{0}dv_{0}du_{1}dv_{1} \Phi\left(\varphi_{uv}(t_{1});\,\varphi'_{uv}(0),\,\varphi'_{uv}(t_{1}),\,\varphi'_{uv}(1)\right)\,\mathcal {E}_{\sigma\sqrt{t_{1}}}(u_{0},\,v_{0})\,\mathcal {E}_{\sigma\sqrt{1-t_{1}}}(u_{1},\,v_{1})\,.
\end{equation}

Instead of the variables $(u_{0},\,v_{0},\,u_{1},\,v_{1})\,,$ it is convenient to use the variables
$$
z_{1}=\varphi(t_{1})\,,\ \ \ x_{0}= \varphi'(0)\,,\ \ \ y_{1}=\varphi'(t_{1})\,,\ \ \ x_{1}=\varphi'(1)\,.
$$
Eqs. (\ref{FiT1}) and (\ref{DerFi01}) lead to
$$
u_{0}=\frac{t_{1}}{z_{1}}\,x_{0},\ \ \ v_{0}= \frac{t_{1}}{z_{1}}\,y_{1},\ \ \ u_{1}=\frac{1-t_{1}}{1-z_{1}}\,y_{1},\ \ \ v_{1}=\frac{1-t_{1}}{1-z_{1}}\,x_{1},
$$
 and
$$
\left|\det \frac{\partial(u_{1},\,v_{1},\,u_{2},\,v_{2})}{\partial (z,\,x_{0},\,x_{1},\,y_{1})} \right|=\frac{\left[t_{1}(1-t_{1})\right]^{2}}{\left[z_{1}(1-z_{1})\right]^{3}}\,y_{1}\,.
$$

Thus, in this case, we get the transparent rule of the functional integration
$$
\int\limits_{Diff^{1}_{+} ([0,\,1]) }\,\Phi\left(\varphi(t_{1});\,\varphi'(0),\,\varphi'(t_{1}),\,\varphi'(1)\right)\ \mu_{\sigma}(d\varphi)
$$
$$
=\left[t_{1}(1-t_{1})\right]^{2}\,\int\limits_{0}^{1}\left[z_{1}(1-z_{1})\right]^{-3}\,dz_{1}\,\int\limits_{0}^{+\infty}\,dx_{0}\,\int\limits_{0}^{+\infty}\,
y_{1}\,dy_{1}\,
\int\limits_{0}^{+\infty}\,dx_{1}
$$
\begin{equation}
   \label{Rule1}
\times\,\Phi\left(z_{1};\,x_{0},\,y_{1},\,x_{1}\right)\,\mathcal {E}_{\sigma\sqrt{t_{1}}}\left(\frac{t_{1}}{z_{1}}\,x_{0},\,\frac{t_{1}}{z_{1}}\,y_{1}\right)\,\mathcal {E}_{\sigma\sqrt{1-t_{1}}}\left(\frac{1-t_{1}}{1-z_{1}}\,y_{1},\,\frac{1-t_{1}}{1-z_{1}}\,x_{1}\right)\,.
\end{equation}

If the integrand $\Phi $ depends on values of the function $\varphi $ and its derivative $\varphi' $ at other points $t_{1},\,t_{2},\,...\,t_{k} $ of the interval, we can continue
 the described procedure
$$
\varphi\rightarrow(\psi_{0},\,\psi_{1})\,,\ \ \psi_{1}\rightarrow(\psi_{2},\,\psi_{3})\,,\ \ \psi_{3}\rightarrow(\psi_{4},\,\psi_{5})\,,\ ...\
\psi_{2k-3}\rightarrow(\psi_{2(k-1)},\,\psi_{2k-1})\,,
$$
and obtain the generalization of (\ref{Rule1})\footnote{In this case, the Jacobian of the substitution is
$$
\left|\det \frac{\partial(u_{1},\,v_{1},\,...\,u_{k+1},\,v_{k+1})}{\partial (z_{1},\,...\,z_{k},\,x_{0},\,y_{1},\,...\,y_{k},\,x_{1})} \right|=\left[t_{1}\cdot\cdot\cdot(1-t_{k})\right]^{2}\,\left[z_{1}\cdot\cdot\cdot(1-z_{k})\right]^{-3}\,y_{1}\cdot\cdot\cdot y_{k}\,.
$$  }
$$
\int\limits_{Diff^{1}_{+} ([0,\,1]) }\,\Phi\left(\varphi(t_{1}),\,...\,\varphi(t_{k});\,\varphi'(0),\,\varphi'(t_{1}),\,...\,\varphi'(t_{k}),\,\varphi'(1)\right)\ \mu_{\sigma}(d\varphi)
$$
$$
=\left[t_{1}(t_{2}-t_{1})\cdot\cdot\cdot(t_{k}-t_{k-1})(1-t_{k})\right]^{2}\,\int\limits_{0}^{1}dz_{1}...\int\limits_{z_{k-1}}^{1}dz_{k}\,
\left[z_{1}(z_{2}-z_{1})\cdot\cdot\cdot(z_{k}-z_{k-1})(1-z_{k})\right]^{-3}
$$
$$
\times\int\limits_{0}^{+\infty}dx_{0}\int\limits_{0}^{+\infty}
y_{1}\,dy_{1}\cdot\cdot\cdot\int\limits_{0}^{+\infty}
y_{k}\,dy_{k}
\int\limits_{0}^{+\infty}dx_{1}\,\Phi\left(z_{1},\,...\,z_{k};\,x_{0},\,y_{1},\,...\,y_{k},\,x_{1}\right)\,\mathcal {E}_{\sigma\sqrt{t_{1}}}\left(\frac{t_{1}}{z_{1}}\,x_{0},\,\frac{t_{1}}{z_{1}}\,y_{1}\right)
$$
$$
\times\,\mathcal {E}_{\sigma\sqrt{t_{2}-t_{1}}}\left(\frac{t_{2}-t_{1}}{z_{2}-z_{1}}\,y_{1},\,\frac{t_{2}-t_{1}}{z_{2}-z_{1}}\,y_{2}\right)\,\cdot\cdot\cdot\,
\mathcal {E}_{\sigma\sqrt{t_{k}-t_{k-1}}}\left(\frac{t_{k}-t_{k-1}}{z_{k}-z_{k-1}}\,y_{k-1},\,\frac{t_{k}-t_{k-1}}{z_{k}-z_{k-1}}\,y_{k}\right)
$$
\begin{equation}
   \label{RuleK}
\times\,\mathcal {E}_{\sigma\sqrt{1-t_{k}}}\left(\frac{1-t_{k}}{1-z_{k}}\,y_{k},\,\frac{1-t_{k}}{1-z_{k}}\,x_{1}\right)\,.
\end{equation}

Note that (\ref{E}) is the only one functional integral we need to evaluate.

In \cite{(BShCorrel)}, we performed the functional integration explicitly and reduced
the functional integral (\ref{E}) to the ordinary integral:
$$
\mathcal {E}_{\sigma }\left(u,\,v\right)
=\left(\frac{2}{\pi\sigma^{2}}\right)^{\frac{3}{2}}\,\frac{1}{\sqrt{uv}}\,\exp\left\{\frac{2}{\sigma^{2}}\left(\pi^{2}-u-v\right) \right\}
$$
\begin{equation}
   \label{Eint}
\times\int\limits_{0}^{+\infty}\,\exp\left\{-\frac{2}{\sigma^{2}}\left(2\,\sqrt{uv}\,\cosh\theta+\theta^{2}\right) \right\}\,\sin\left(\frac{4\pi\theta}{\sigma^{2}} \right)\,\sinh(\theta)\,d\theta \,.
\end{equation}

Now the functional integrals (\ref{Rule1}) and (\ref{RuleK}) are expressed as ordinary multiple integrals.

Here, we split the interval at the points the integrand depends on. However the following question may arise. When we split the interval at an arbitrary
point, does the dependence on the point appear in the result according to (\ref{RuleK}), even if the integrand does not depend on the point?
To prove that the dependence is in fact fictitious, we apply (\ref{Rule1}) to the function $\mathcal {E}(u,\,v) $ and obtain the following convolution rule:
$$
\mathcal {E}_{\sigma}(u,\,v)=\left[t_{\ast}(1-t_{\ast})\right]^{2}\,\int\limits_{0}^{1}\left[z_{\ast}(1-z_{\ast})\right]^{-3}\,dz_{\ast}\,\int\limits_{0}^{+\infty}\,
y_{\ast}\,dy_{\ast}\,
$$
\begin{equation}
   \label{ConvRule}
\times\,\mathcal {E}_{\sigma\sqrt{t_{\ast}}}\left(\frac{t_{\ast}}{z_{\ast}}\,u,\,\frac{t_{\ast}}{z_{\ast}}\,y_{\ast}\right)\,\mathcal {E}_{\sigma\sqrt{1-t_{\ast}}}\left(\frac{1-t_{\ast}}{1-z_{\ast}}\,y_{\ast},\,\frac{1-t_{\ast}}{1-z_{\ast}}\,v\right)\,.
\end{equation}
where $t_{\ast}$ is an arbitrary point of the interval $[0,\,1]\,.$

\section{Correlation functions defined as functional integrals over the group $Diff^{1}_{+}(\textbf{R}) $ }
\label{sec:diffR}

Consider functional integrals over $Diff^{1}_{+}(\textbf{R}) $ of the form
\begin{equation}
   \label{NewFI}
\int \limits_{Diff^{1}_{+}(\textbf{R}) }\,\Psi \left(f(\tau_{1}),\,f(\tau_{2}),\,... \right)\
 \exp\left\{\frac{1}{\sigma^{2}}\int \limits _{-\infty}^{+\infty}\,\mathcal{S}ch\{f,\,\tau\}\,d\tau\, \right\}\,df
\end{equation}

The naive direct functional integration in (\ref{NewFI}) with  $SL(2, \textbf{R})-$invariant integrands $\Psi $ leads to senseless infinite results.

Although $SL(2, \textbf{R})$ is not a subgroup of the group  $Diff^{1}_{+}(\textbf{R})$\footnote{
For example, the transformation $ g_{a}\circ f=g_{a} (f) = (f+a)^{-1}\,\  \left(g_{a}\in SL(2, \textbf{R})\,,\ f\in Diff^{1}_{+}(\textbf{R})\right)$ removes $ g_{a}\circ f $ from $Diff^{1}_{+}(\textbf{R})\,.$},
 the group $ P=A\,N\ $  $\left(P\subset SL(2, \textbf{R})\right)$ consisting of transformations $f\rightarrow af+b$ (see \cite{(Lang)}, ch. III, par. 1) is a
subgroup of  $Diff^{1}_{+}(\textbf{R})\,.$
Since the group $P$ is noncompact, to obtain a finite result one should integrate over the quotient space  $Diff^{1}_{+}(\textbf{R})/ P\,.$

One can factorize the integration space and exclude the non-relevant degrees of freedom in different ways. For example, one can fix
the "gauge" putting\footnote{Exactly the same gauge-fixing conditions were supposed in \cite{(BAK1)}.}
$$
f(0)=0\,, \ \ f'(0)=1\,.
$$

Now we consider the two-point correlation function
\begin{equation}
   \label{Kam2}
\mathcal {G}_{2}^{n}(0,\,\tau)=\int \limits_{Diff^{1}_{+}(\textbf{R})/P}\,\left(\mathcal{O}_{f}\left(0\,,\,\tau\right)\right)^{n}\
 \exp\left\{\frac{1}{\sigma^{2}}\int \limits _{-\infty}^{+\infty}\,\mathcal{S}ch\{f,\,t\}\,dt\, \right\}\,df\,.
\end{equation}

After the substitution $f(\tau)=\int\limits^{\tau}_{0}\exp\{\xi(t)\}dt\,,\ \ \xi(0)=0$,  it looks like
$$
\mathcal {G}_{2}^{n}(0,\,\tau)=\int \limits_{C((-\infty,\,+\infty))}\,\frac{e^{\frac{n}{4}(\xi(\tau))}}{\left(\int \limits_{0}^{\tau}\,e^{\xi(t)}dt\right)^{\frac{n}{2}}}
 \exp\left\{-\frac{1}{2\sigma^{2}}\int \limits _{-\infty}^{+\infty}\,\left(\xi'(t)\right)^{2}\,dt\, \right\}\,d\xi\,.
$$
Due to the Markov property of the Wiener measure, it can be rewritten as\footnote{Cf. equation (18) in \cite{(BAK1)}.}
\begin{equation}
   \label{Bksi}
\mathcal {G}_{2}^{n}(0,\,\tau)=\int \limits_{C([0,\,\tau])}\,\frac{e^{\frac{n}{4}(\xi(\tau))}}{\left(\int \limits_{0}^{\tau}\,e^{\xi(t)}dt\right)^{\frac{n}{2}}}
 \exp\left\{-\frac{1}{2\sigma^{2}}\int \limits_{0}^{\tau}\,\left(\xi'(t)\right)^{2}\,dt\, \right\}\,d\xi\,.
\end{equation}

Instead of transforming the Wiener integral (\ref{Bksi}) to the functional integral in Liouville theory as it was proposed in \cite{(BAK1)}, \cite{(BAK2)},  we re-scale the time variable $t=\tau\,\bar{t}$ and substitute $\eta(\bar{t})=\xi(t)\,. $
Then (\ref{Bksi}) has the form
\begin{equation}
   \label{Beta}
\mathcal {G}_{2}^{n}(0,\,\tau)=\tau^{-\frac{n}{2}}\,\int \limits_{C([0,\,1])}\,\frac{e^{\frac{n}{4}\eta(1)}}{\left(\int \limits_{0}^{1}\,e^{\eta(\bar{t})}d\bar{t}\right)^{\frac{n}{2}}}
 \exp\left\{-\frac{1}{2\sigma^{2}\tau}\int \limits_{0}^{1}\,\left(\eta'(\bar{t})\right)^{2}\,d\bar{t}\, \right\}\,d\eta\,.
\end{equation}

In terms of the function
$$
\varphi(t)=\frac{\int \limits _{0}^{t}\,e^{\eta(\bar{t})}d\bar{t}}{\int \limits _{0}^{1}\,e^{\eta(\bar{t})}d\bar{t} }\,,
$$
the integral (\ref{Beta}) is written as
\begin{equation}
   \label{BintDiff}
\tau^{-\frac{n}{2}}\,\int \limits_{Diff^{1}_{+}([0,\,1])}\,\left(\varphi'(0) \varphi'(1) \right)^{\frac{n}{4}}\,\mu _{\sigma\sqrt{\tau}}(d\varphi)
=\tau^{-\frac{n}{2}}\,\int \limits_{0}^{+\infty}\,\int \limits_{0}^{+\infty}\,\left(uv \right)^{\frac{n}{4}}\,\mathcal {E} _{\sigma\sqrt{\tau}}(u\,,v)\,du\,dv\,.
\end{equation}

Now we use the explicit form of the function $\mathcal {E}$ (\ref{Eint}). After the substitution
$$
u=\rho\cos^{2}\omega\,,\ \ \ \ v=\rho\sin^{2}\omega\,,\ \ \ \ \xi=\tan \omega \,,
$$
we can integrate with the help of the table of integrals \cite{(Prud)}.
The result is
$$
\mathcal {G}_{2}^{n}(0,\,\tau)=2\pi^{-\frac{3}{2}}\left( \frac{\sigma^{2}}{2}\right)^{\frac{n-1}{2}}\,B\left(\frac{n+1}{2},\, \frac{n+1}{2}\right)\,\tau^{-\frac{1}{2}}\,\int\limits_{0}^{+\infty}\,
d\theta\,\sin\left(\frac{4\pi\,\theta}{\sigma^{2}\tau}\right)
\sinh \theta
$$
\begin{equation}
\label{gnResult}
\times \exp\left\{-\frac{2}{\sigma^{2}\tau}\left[
\theta^{2}-\pi^{2}\right]\right\}\
{ }_{2}F_{1}\left(\frac{n+1}{2},\, \frac{n+1}{2};\,\frac{n+1}{2}+1;\,\frac{1-\cosh \theta}{2}\right)\,.
\end{equation}

In the case $n=2$,  it looks like
\begin{equation}
\label{g2Result}
\mathcal {G}_{2}^{2}(0,\,\tau)=\left( \frac{\sigma^{2}}{2\pi^{3}}\right)^{\frac{1}{2}}\tau^{-\frac{1}{2}}\int\limits_{0}^{+\infty}
d\theta\exp\left\{-\frac{2}{\sigma^{2}\tau}\left[
\theta^{2}-\pi^{2}\right]\right\}
\sin\left(\frac{4\pi\theta}{\sigma^{2}\tau}\right)
\left[\frac{\theta \cosh\theta }{\sinh^{2}\theta }-\frac{1}{\sinh\theta }\right].
\end{equation}

The asymptotic forms of $\mathcal {G}_{2}^{n}(0,\,\tau)$ at $\tau\rightarrow\infty$
$$
\left(\mathcal {G}_{2}^{n}(0,\,\tau)\right)^{As}_{\infty}= 4\pi^{-\frac{1}{2}}\left( \frac{\sigma^{2}}{2}\right)^{\frac{n-3}{2}}\,B\left(\frac{n+1}{2},\, \frac{n+1}{2}\right)
$$
\begin{equation}
\label{gnAs}
\times\,\int\limits_{0}^{+\infty}
\,
  { }_{2}F_{1}\left(\frac{n+1}{2},\, \frac{n+1}{2};\,\frac{n+1}{2}+1;\,\frac{1-\cosh \theta}{2}\right)\,\theta\,
\sinh \theta \,d\theta\ \times\ \tau^{-\frac{3}{2}}\,,
\end{equation}
and at $\tau\rightarrow 0$
\begin{equation}
\label{gnAszero}
\left(\mathcal {G}_{2}^{n}(0,\,\tau)\right)^{As}_{0}=const\,\times\,\tau^{-\frac{n}{2}}\,,
\end{equation}
immediately follow from (\ref{gnResult}).

In the same way, we can explicitly evaluate functional integrals for other correlation functions in this theory.
In particular, for
$\mathcal {G}_{4}^{TO}(0,\,\tau_{_{1}};\,\tau_{2},\,\tau_{3})$
 we have $(0<\tau_{1}<\tau_{2}<\tau_{3})$
$$
\mathcal {G}_{4}^{TO}(0,\,\tau_{_{1}};\,\tau_{2},\,\tau_{3})=\int \limits_{Diff^{1}_{+}(\textbf{R})/P }\,\mathcal{O}_{f}\left(0\,,\,\tau_{1}\right)
\mathcal{O}_{f}\left(\tau_{2}\,,\,\tau_{3}\right)
\,
 \exp\left\{\frac{1}{\sigma^{2}}\int \limits _{-\infty}^{+\infty}\,\mathcal{S}ch\{f\,,\,t\}\,dt\, \right\}\,df
$$
\begin{equation}
\label{gTO}
=\mathcal {G}_{2}(0,\,\tau_{_{1}})\,\mathcal {G}_{2}(\,\tau_{2},\,\tau_{3})\,.
\end{equation}

The OTO four-point correlation function is given by $(0<\tau_{1}<\tau_{2}<\tau_{3})$
$$
\mathcal {G}_{4}^{OTO}(0,\,\tau_{_{2}};\,\tau_{1},\,\tau_{3})=\int \limits_{Diff^{1}_{+}(\textbf{R})/P }\,\mathcal{O}_{f}\left(0\,,\,\tau_{2}\right)
\mathcal{O}_{f}\left(\tau_{1}\,,\,\tau_{3}\right)
\,
 \exp\left\{\frac{1}{\sigma^{2}}\int \limits _{-\infty}^{+\infty}\,\mathcal{S}ch\{f\,,\,t\}\,dt\, \right\}\,df
$$
$$
=\pi^{\frac{9}{2}}\left(\frac{\sigma^{2}}{2}\right)^{\frac{1}{2}}\left[\tau_{1}(\tau_{2}-\tau_{1})(\tau_{3}-\tau_{2})\right]^{-\frac{1}{2}}
\int\limits_{0}^{1}dz_{1}\int\limits_{z_{1}}^{1}dz_{2}\,
\left[z_{1}(z_{2}-z_{1})(1-z_{2})\right]^{-2}\left[z_{2}(1-z_{1}) \right]^{-\frac{1}{2}}
$$
$$
\times\int\limits_{0}^{+\infty}\,\int\limits_{0}^{+\infty}\,\int\limits_{0}^{+\infty}d\theta_{1}\, d\theta_{2}\,d\theta_{3}\,\exp\left\{-\frac{2}{\sigma^{2}}\left[\frac{\theta_{1}^{2}-\pi^{2}}{\tau_{1}}+
\frac{\theta_{2}^{2}-\pi^{2}}{(\tau_{2}-\tau_{1})}+
\frac{\theta_{3}^{2}-\pi^{2}}{(\tau_{3}-\tau_{2})}\right]\right\}
$$
$$
\times\,
\sin\left(\frac{4\pi\,\theta_{1}}{\sigma^{2}\,\tau_{1}}\right)\,
\sin\left(\frac{4\pi\,\theta_{2}}{\sigma^{2}\,(\tau_{2}-\tau_{1})}\right)\,\sin\left(\frac{4\pi\,\theta_{3}}{\sigma^{2}\,(\tau_{3}-\tau_{2})}\right)\, \sinh\theta_{1}\,\sinh\theta_{2}\,\sinh\theta_{3}
$$
$$
\times\int\limits_{0}^{+\infty}dx_{0}\int\limits_{0}^{+\infty}
dy_{1}\int\limits_{0}^{+\infty}
dy_{2}\int\limits_{0}^{+\infty}
dx_{1}\left[x_{0}x_{1}\right]^{-\frac{1}{4}}  \left[y_{1}y_{2}\right]^{\frac{1}{4}}\exp\left\{
-\frac{1}{z_{1}}\left[x_{0}+y_{1}+2\sqrt{x_{0}y_{1}}\cosh \theta_{1} \right]\right\}
$$
\begin{equation}
\label{gOTO}
\times \exp\left\{
-\frac{1}{z_{2}-z_{1}}\left[y_{1}+y_{2}+2\sqrt{y_{1}y_{2}}\cosh \theta_{2} \right]
-\frac{1}{1-z_{2}}\left[y_{2}+x_{1}+2\sqrt{y_{2}x_{1}}\cosh \theta_{3} \right] \right\}\,.
\end{equation}

Evaluated functional integrals (\ref{Kam2}), (\ref{Bksi}), and (\ref{gTO}), (\ref{gOTO}) for $\mathcal {G}_{2}$, and for $\mathcal {G}_{4}^{TO\,;\,OTO}$  (including  gauge-fixing conditions) are precisely the same as those in \cite{(BAK1)}. The results of calculation in \cite{(BAK1)} and in the present paper obtained in the  different ways have  exactly the same behaviour.
Thus we reproduce the results for correlation functions obtained in \cite{(BAK1)},  although in a slightly different form.

\section{Quasi-invariance of the measure as a key to functional integration}
\label{sec:quasi-inv}

As an important step in evaluating functional integrals over the group of diffeomorphisms of the circle (\ref{GenFormI}), we consider functional integrals
over the group of diffeomorphisms of the interval of the form
\begin{equation}
   \label{FI3A}
J^{(\alpha)}\equiv\int \limits _{Diff_{+}^{1}([0, 1])}\,\Psi(\phi,\,\phi')\,
 \exp\left\{\frac{2\alpha^{2}}{\sigma^{2}}\int \limits _{0}^{1}\left(\phi' (t)\right)^{2}dt  \right\}\, \mu_{\sigma} (d\phi)\,.
\end{equation}

The quasi-invariance of the measure  (\ref{Measure}) with respect to the left action of the subgroup  $Diff^{3}_{+}([0,\,1])$
$\,(\varphi=g\circ \phi=g( \phi)\,,\ \ g\in Diff^{3}_{+}([0,\,1])\,,\ \ \varphi\,, \phi\in Diff^{1}_{+}([0,\,1])\,)\,$ is written  as\footnote{The form of Radon - Nikodim derivative is related to that of the corresponding Wiener measure
\cite{(Shepp)}.} \cite{(Shavgulidze1978)}-\cite{(Shavgulidze2000)}
  $$
  \int \limits _{Diff^{1}_{+}([0, 1])}\,\Phi(\varphi,\,\varphi')\,\mu_{\sigma}(d\varphi)=\frac{1}{\sqrt{g'(0)g'(1)}}\int \limits _{Diff^{1}_{+}([0, 1])}\,\Phi\left(g(\phi),\,(g(\phi))'\right)
  $$
  \begin{equation}
   \label{FI1}
  \times \exp\left\{ \frac{1}{\sigma^{2}}\left[   \frac{g''(0)}{g'(0)}\phi'(0)-  \frac{g''(1)}{g'(1)}\phi'(1)\right]   +        \frac{1}{\sigma^{2}}\int \limits _{0}^{1}\, \mathcal{S}ch\left\{g,\,\phi(t)\right\}\,\left(\phi' (t)\right)^{2}\,dt  \right\} \,\mu_{\sigma}(d\phi)\,.
\end{equation}

Let the function $g$  be
\begin{equation}
   \label{g}
g(t)=g_{\alpha}(t)=\frac{1}{2}\left[ \frac{1}{\tan\frac{\alpha}{2}}\tan\left(\alpha(t-\frac{1}{2}) \right)+1 \right]\,.
\end{equation}
In this case,
\begin{equation}
   \label{tan}
g'_{\alpha}(0)=g'_{\alpha}(1)=\frac{\alpha}{\sin \alpha}\,,
\ \ \ -\frac{g''_{\alpha}(0)}{g'_{\alpha}(0)}=\frac{g''_{\alpha}(1)}{g'_{\alpha}(1)}=2\alpha\tan\frac{\alpha}{2}\,, \ \ \ \mathcal{S}ch\{g_{\alpha}\,,\,t\}=2\alpha^{2}\,,
\end{equation}
and the equation (\ref{FI1}) looks like:
$$
\int \limits _{Diff_{+}^{1}([0, 1])}\Phi(\varphi,\,\varphi')\,\mu_{\sigma}(d\varphi)=\frac{\sin \alpha}{\alpha}\,\int \limits _{Diff_{+}^{1}([0, 1])}\Phi\left(g_{\alpha}(\phi),\,(g_{\alpha}(\phi))'\right)
 $$
\begin{equation}
   \label{FI2}
\times\,\exp\left\{-\frac{2\alpha}{\sigma^{2}}\tan\frac{\alpha}{2}\left(\phi'(0)+
\phi'(1)\right)\right\}\,\exp\left\{\frac{2\alpha^{2}}{\sigma^{2}}\int \limits _{0}^{1}\left(\phi' (t)\right)^{2}dt  \right\} \mu_{\sigma} (d\phi)\,.
\end{equation}

Denote
\begin{equation}
   \label{FPsiI}
\Psi (\phi,\,\phi')=\exp\left\{-\frac{2\alpha}{\sigma^{2}}\tan\frac{\alpha}{2}\left(\phi'(0)+\phi'(1)\right)\right\}\,\Phi\left(g_{\alpha}(\phi),\,
(g_{\alpha}(\phi))'\right)\,,
\end{equation}
then
\begin{equation}
   \label{FPsi}
\Phi(\varphi,\,\varphi')=\exp\left\{\frac{2\alpha}{\sigma^{2}}\tan\frac{\alpha}{2}\left(\left(g^{-1}_{\alpha}(\varphi)\right)'(0)
+\left(g^{-1}_{\alpha}(\varphi)\right)'(1)\right)\right\}\,\Psi \left(g^{-1}_{\alpha}(\varphi),\,\left(g^{-1}_{\alpha}(\varphi)\right)'\right)\,.
\end{equation}
Now the equation (\ref{FI2}) has the form
$$
\frac{\alpha}{\sin \alpha}\,\int \limits _{Diff_{+}^{1}([0, 1])}\Phi(\varphi,\,\varphi')\,\mu_{\sigma}(d\varphi)=\int \limits _{Diff_{+}^{1}([0, 1])}\,\Psi(\phi,\,\phi')\,
 \exp\left\{\frac{2\alpha^{2}}{\sigma^{2}}\int \limits _{0}^{1}\left(\phi' (t)\right)^{2}dt  \right\}\, \mu_{\sigma} (d\phi)\,.
$$

For the function $g_{\alpha} $ given by (\ref{g}), the inverse function is
$$
\left(g_{\alpha}^{-1}(\varphi)\right)(t)=\frac{1}{\alpha}\arctan\left[\tan \frac{\alpha}{2}\,\left(2\varphi(t)-1\right)\right]+\frac{1}{2}\,,
$$
$$
\left(g^{-1}_{\alpha}(\varphi)\right)'(0)=\frac{\sin \alpha}{\alpha}\,\varphi'(0)\,,\ \ \ \ \ \left(g^{-1}_{\alpha}(\varphi)\right)'(1)=\frac{\sin \alpha}{\alpha}\,\varphi'(1)\,,
$$
and
$$
\exp\left\{\frac{2\alpha}{\sigma^{2}}\tan\frac{\alpha}{2}\left(\left(g^{-1}_{\alpha}(\varphi)\right)'(0)
+\left(g^{-1}_{\alpha}(\varphi)\right)'(1)\right)\right\}=\exp\left\{\frac{4\,\sin ^{2}\frac{\alpha}{2}}{\sigma^{2}}\left(\varphi'(0)+\varphi'(1)\right)\right\}\,.
$$

Thus the integral (\ref{FI3A}) is transformed into
\begin{equation}
   \label{FIJ}
J^{(\alpha)}=\frac{\alpha}{\sin \alpha}
\int \limits _{Diff_{+}^{1}([0, 1])}\Psi \left(g^{-1}_{\alpha}(\varphi),\,\left(g^{-1}_{\alpha}(\varphi)\right)'\right)\,\exp\left\{\frac{4}{\sigma^{2}}\sin^{2}\frac{\alpha}{2}\left(\varphi'(0)
+\varphi'(1)\right)\right\}\mu_{\sigma} (d\varphi)\,.
\end{equation}

Note that generally, functional integrals of the form (\ref{FIJ}) diverge at $ \alpha = \pi  $ indicating that functional integrals (\ref{GenFormI}) also diverge.

\section{How to exclude the infinite input of $SL(2,\textbf{R})$$SL(2,\textbf{R})$}
\label{sec:ren}

The reason of the divergence of functional integrals (\ref{GenFormI})
is  in the invariance
\begin{equation}
   \label{invariance}
\mathcal{S}ch \{\varphi,\,t\}+2\pi^{2}\left(\varphi'(t)\right)^{2}=\mathcal{S}ch \{\psi,\,t\}+2\pi^{2}\left(\psi'(t)\right)^{2}
\end{equation}
under the left action of $SL(2,\textbf{R})\,:$ $\,\varphi(t)=\gamma(\psi(t))\,\ \left(\,\gamma\in SL(2,\textbf{R})\,\ \  \varphi,\,\psi \in Diff^{1}_{+}(S^{1})\,\right)\,.$

To prove the invariance (\ref{invariance})
 consider the following realization of the group $SL(2,\textbf{R})$
\begin{equation}
   \label{phi}
  \gamma (\psi)=\frac{1}{i2\pi}\,\log\,\frac{e^{i2\pi \psi}+z}{\bar{z}e^{i2\pi \psi}+1}\ ,
\end{equation}
and use the known property of Schwarzian derivative
\begin{equation}
   \label{property}
   \mathcal{S}ch\{\gamma\circ \psi,\,t\}=\mathcal{S}ch\, \{\gamma,\,\psi(t)\}\,\left(\psi'(t)\right)^{2}+\mathcal{S}ch\{\psi,\,t\}\,.
\end{equation}

One can exclude the non-relevant $SL(2,\textbf{R})$ degrees of freedom by fixing three parameters of the group. It may create some difficulties for subsequent calculations (see, e.g., \cite{(SW)} p.9, footnote 7), since  $SL(2,\textbf{R})$ gauge-fixing conditions cannot be imposed at one point of the unit circle only, but inevitably contain  values of the function $\varphi $ or of its first derivative  at some other points of the circle.
However, using the method of section \ref{sec:red}, we can overcome the difficulties easily.

Moreover, we can factorize  the measure on $Diff^{1}_{+}(S^{1}) $ (see also \cite{(BShFactor)})
as
\begin{equation}
   \label{FactMeasure}
   \tilde{\mu}_{\sigma}(d\varphi)=\nu _{H}(d\gamma)\ \tilde{\mu}_{\sigma}^{X}(d\psi) \,,
\end{equation}
where
\begin{equation}
   \label{MuTilde}
   \tilde{\mu}_{\sigma}(d\varphi)\equiv \exp\left\{\frac{2\pi^{2}}{\sigma^{2}}\int \limits _{S^{1}}\left(\varphi' (t)\right)^{2}dt  \right\}\, \mu_{\sigma} (d\varphi)\,,
\end{equation}
and $\tilde{\mu}_{\sigma}^{X}(d\psi)$ is the corresponding measure on the quotient space $Diff^{1}_{+}(S^{1})/SL(2,\textbf{R})\,. $

To realize the factorization (\ref{FactMeasure}),  we take 3 points on the circle $t_{k}=\frac{k}{3}\,,\ \ k=0,\,1,\,2\,, $ and rewrite the integral
\begin{equation}
   \label{FIpsi}
\frac{1}{\sqrt{2\pi}\sigma}\int \limits _{Diff_{+}^{1}(S^{1})}\,\Psi(\varphi)\,
 \exp\left\{\frac{2\pi^{2}}{\sigma^{2}}\int \limits _{S^{1}}\left(\varphi' (t)\right)^{2}dt  \right\}\, \mu_{\sigma} (d\varphi)
\end{equation}
as
\begin{equation}
   \label{FI3tau}
\frac{1}{\sqrt{2\pi}\sigma}\,\int\,d\tau_{0}\,d\tau_{1}\,d\tau_{2}\int\limits_{Diff^{1}_{+} (S^{1}) }\Psi(\varphi)\prod\limits_{k=0,\,1,\,2}\delta\left(\varphi(t_{k})-\tau_{k} \right)\,
 \tilde{ \mu}_{\sigma} (d\varphi)\,.
\end{equation}

Then
we fix the 3 parameters of $SL(2,\textbf{R})$ assuming $\gamma_{\tau}(t_{k})=\tau_{k}\,,$ with $\tau_{k}$ being fixed.
Using the equality
$$
\delta\left(\gamma_{\tau}(\psi(t_{k}))-\tau_{k} \right)=\frac{1}{\gamma\,'_{\tau}(t_{k})}\delta\left(\psi(t_{k})-t_{k} \right)\,,
$$
we transform the integral (\ref{FI3tau}) into the form
\begin{equation}
   \label{FI3f}
\frac{1}{\sqrt{2\pi}\sigma}\int\prod\limits_{i=0,\,1,\,2}\frac{d\tau_{i}}{\gamma\,'_{\tau}(t_{i})}\int\limits_{Diff^{1}_{+} (S^{1});\, SL(2,R) gauge fixed }\Psi(\gamma_{\tau}\circ\ \psi)\prod\limits_{k=0,\,1,\,2}\delta\left(\psi(t_{k})-t_{k} \right)\,
 \tilde{ \mu}_{\sigma} (d\psi)\,.
\end{equation}

It is convenient to consider the following realization of $SL(2,\textbf{R})$
\begin{equation}
   \label{SLrealiz}
\gamma_{\tau}(t)=\frac{1}{\pi}arc\cot\left\{A\cot(\pi t-\pi\theta)+B \right\}
\end{equation}
with the 3 parameters $A,\ B,\ \theta  $ related to $\tau_{0},\,\tau_{1},\,\tau_{2}\,.$

In terms of the  parameters $A,\ B, \ \theta  $, the measure $\prod\limits_{i=0,\,1,\,2}\frac{d\tau_{i}}{\gamma \,'_{\tau}(t_{i})} $ in (\ref{FI3f}) is written as
$$
const\, \frac{dB\,dA}{A^{2}}\,d\theta
$$
that is the Haar measure of $SL(2,\textbf{R})\ \ \nu_{H}(d\gamma)$ \cite{(Lang)}.

For $SL(2,\textbf{R})$ invariant integrands $\Psi(\gamma_{\tau}\circ\, \psi)=\Psi( \psi)\,, $ the integral
\begin{equation}
   \label{FI3AS}
\int \limits _{Diff_{+}^{1}(S^{1})}\,\Psi(\varphi)\,
 \exp\left\{\frac{2\pi^{2}}{\sigma^{2}}\int \limits _{S^{1}}\left(\varphi' (t)\right)^{2}dt  \right\}\, \mu_{\sigma} (d\varphi)
\end{equation}
 is factorized
\begin{equation}
   \label{FI3f2}
\frac{1}{\sqrt{2\pi}\sigma}\int\limits_{ SL(2,R) }\nu_{H}(d\gamma)\,\int\limits_{Diff^{1}_{+} (S^{1})/ SL(2,R)}\Psi( \psi)\prod\limits_{k=0,\,1,\,2}\delta\left(\psi(t_{k})-t_{k} \right)\,
 \tilde{ \mu}_{\sigma} (d\psi)\,.
\end{equation}

Therefore we demonstrate that
$Diff^{1}_{+}(S^{1}) $ , as the space to integrate over, is equivalent to the Cartesian product
\begin{equation}
   \label{FactSpace}
Diff^{1}_{+}(S^{1})\cong\,SL(2,\textbf{R})\ \times\ Diff^{1}_{+}(S^{1})/SL(2,\textbf{R})\,.
\end{equation}

Thus we prove the correctness of the renormalization procedure proposed in \cite{(BShCorrel)}. It
factors out the infinite input of the noncompact group $SL(2,\textbf{R})$, and leads to the finite results  that are
  functional integrals over the quotient space $Diff^{1}_{+}(S^{1})/SL(2,\textbf{R})\,. $

The procedure has the following main steps.
First, one represents functional integrals over the group  $Diff^{1}_{+}(S^{1})$ as integrals over the group  $Diff^{1}_{+}([0, 1])$ with the ends of the interval glued (\ref{Equality}). Then,
to regularize the functional integral, one substitutes $\alpha$ for $\pi$ in the exponent in the integrand representing the integral in the form (\ref{FI3A}).

The renormalized integral is defined as the limit at $\alpha\rightarrow\pi - 0 $ of the ratio of $ J^{(\alpha)}$ to the corresponding
integral over the group $SL(2, \textbf{R})\,.$
For $SL(2, \textbf{R})$-invariant integrands, the denominator is proportional to
the regularized volume of the group $SL(2,\textbf{R})$  explicitly evaluated in \cite{(BShCorrel)}:
$$
V^{(\alpha)}_{SL(2, \textbf{R})}=\int \limits _{SL(2,\textbf{R})}\exp \left\{-\frac{2\left[ \pi^{2}-\alpha^{2}\right]}{\sigma^{2}}\,\int \limits _{0}^{1}(\gamma'(t))^{2}dt  \right\}  \nu_{H}(d\gamma)=\frac{\pi\sigma^{2}}{\pi^{2}-\alpha^{2}}\,\exp\left\{-\frac{2\left(\pi^{2} - \alpha^{2} \right)}{\sigma^{2}} \right\}\,,
$$
and the renormalized functional integral is
\begin{equation}
   \label{JRen}
J^{(R)}=\lim \limits_{\alpha\rightarrow\pi - 0}\  \frac{ J^{(\alpha)}}{ V^{(\alpha)}_{SL(2, \textbf{R})}}\,.
\end{equation}

In particular, the two-point correlation function $\tilde{G}_{2}(t_{1},\,t_{2}) $ is given by the functional integral
$$
 \tilde{G}_{2}(t_{1},\,t_{2})=\int \limits_{Diff^{1}_{+}(S^{})}
\mathcal{O}_{\varphi}\left(t_{1}\,,\,t_{2}\right)\,
 \exp\left\{\frac{ 2\pi^{2}}{\sigma^{2}}\int \limits _{S^{1}}\,\left(\varphi'(t)\right)^{2}\,dt\, \right\}\,\mu_{\sigma}(d\varphi)
$$
\begin{equation}
   \label{GavFI}
= \sqrt{2\pi}\sigma\int \limits_{Diff^{1}_{+}([0,\,1])}\delta\left(\frac{\varphi'(1)}{\varphi'(0)}-1 \right)\,
\mathcal{O}_{\varphi}\left(t_{1}\,,\,t_{2}\right)\,
 \exp\left\{\frac{ 2\pi^{2}}{\sigma^{2}}\int \limits _{0}^{1}\,\left(\varphi'(t)\right)^{2}\,dt\, \right\}\,\mu_{\sigma}(d\varphi)\,.
\end{equation}
After the regularization, it is
$$
 \tilde{G}_{2}^{(\alpha )}(t_{1},\,t_{2})= \sqrt{2\pi}\sigma\int \limits_{Diff^{1}_{+}([0,\,1])}\delta\left(\frac{\varphi'(1)}{\varphi'(0)}-1 \right)\,
\mathcal{O}_{\varphi}\left(t_{1}\,,\,t_{2}\right)
 \exp\left\{\frac{ 2\alpha^{2}}{\sigma^{2}}\int \limits _{0}^{1}\,\left(\varphi'(t)\right)^{2}\,dt\, \right\}\mu_{\sigma}(d\varphi)\,.
$$
For the factor in the integrand, we have
$$
\delta\left(\frac{\left(g^{-1}_{\alpha}(\varphi)\right)'(1)}{\left(g^{-1}_{\alpha}(\varphi)\right)'(0)}-1 \right)\,\frac{\left(\left(g^{-1}_{\alpha}(\varphi)\right)'(t_{1})\left(g^{-1}_{\alpha}(\varphi)\right)'(t_{2})\right)^{\frac{1}{4}}}
{\left |\sin \left[\pi \left(g^{-1}_{\alpha}(\varphi)\right) ( t_{2})-\pi\left(g^{-1}_{\alpha}(\varphi)\right)(t_{1})\right]\right|^{\frac{1}{2}}}
$$
$$
=\pi^{-\frac{1}{2}}\,\delta\left(\frac{\varphi'(1)}{\varphi'(0)}-1 \right)\,\frac{\left(\varphi'(t_{1})\varphi'(t_{2})\right)^{\frac{1}{4}}}{\left |\varphi ( t_{2})-\varphi(t_{1})\right|^{\frac{1}{2}}}+O(\pi-\alpha)\,.
$$

Thus the regularized correlation function has the form
$$
  \tilde{G}^{(\alpha)}_{2}(t_{1},\,t_{2})=\sqrt{2}\sigma\,\frac{\alpha}{\sin \alpha}
$$
 \begin{equation}
   \label{2GSYK}
 \times\int \limits_{Diff^{1}_{+}([0,\,1])} \delta\left(\frac{\varphi'(1)}{\varphi'(0)}-1 \right)\, \frac{\left(\varphi'(t_{1})\varphi'(t_{2})\right)^{\frac{1}{4}}}{\left |\varphi ( t_{2})-\varphi(t_{1})\right|^{\frac{1}{2}}}
 \exp\left\{\frac{8}{\sigma^{2}}\varphi'(0) \right\}\,\mu_{\sigma}(d\varphi)+ O(1)\,.
\end{equation}

Since
 the integrand is $SL(2, \textbf{R})$-invariant, the renormalized  correlation function is
\begin{equation}
   \label{GRen}
G_{2}\,\equiv\,\tilde{G}^{(R)}_{2}=\lim \limits_{\alpha\rightarrow\pi - 0}\  \frac{ \tilde{G}^{(\alpha)}_{2}}{ V^{(\alpha)}_{SL(2, \textbf{R})}}\,.
\end{equation}
The singularities $ (\pi - \alpha)^{-1}$ in the nominator and the denominator cancel each other, and we
obtain the finite result for the correlation function.

\section { Correlation functions as renormalized functional integrals over the group  $Diff^{1}_{+}(S^{1}) $}
\label{sec:correl}

The renormalized two-point correlation function\footnote{
Note that the integral (\ref{2GSYK}) is invariant under the shifts of the variable $t$ (that is, under the choice of the point on the $S^{1}$ corresponding to the left end of the interval $[0,\,1]$):
$$
  G_{2}(t_{1},\,t_{2})\,=\,  G_{2}(0,\,t_{2}-t_{1})\,.
$$}
 looks like
$$
G_{2}^{n}\left(0,\,t_{1} \right)
$$
$$
=\frac{2\pi}{\sigma^{2}}\pi^{-\frac{n}{2}}\int\limits_{Diff^{1} ([0,1]) }\delta\left(\frac{\varphi'(1)}{\varphi'(0)}-1 \right)\left(
\frac{\left(\varphi'(t_{1})\varphi'(0)\right)^{\frac{1}{4}}}{\left |\varphi ( t_{1})\right|^{\frac{1}{2}}}\right)^{n}
\exp\left\{\frac{8}{\sigma^{2}}\left(\varphi'(0)\right)\right\}\,\mu_{\sigma}(d\varphi)
$$
$$
= \frac{2\pi}{\sigma^{2}}\pi^{-\frac{n}{2}}\,\left[t_{1}\,(1-t_{1})\right]^{2}\int\limits_{0}^{1}\,dz_{1}\,z_{1}^{-3-\frac{n}{2}}(1-z_{1})^{-3}\, \int\limits_{0}^{+\infty}\int\limits_{0}^{+\infty}dx_{0}dy_{1}\,\left(x_{0}\,y_{1}\right)^{1+\frac{n}{4}}
$$
\begin{equation}
   \label{Ge}
\times\,\exp\left\{\frac{8}{\sigma^{2}}
x_{0} \right\}\mathcal {E}_{\sigma \sqrt{t_{1}}}\left(\frac{t_{1}}{z_{1}}x_{0},\,\frac{t_{1}}{z_{1}}y_{1}\right)\,\mathcal {E}_{\sigma \sqrt{1-t_{1}}}\left( \frac{1-t_{1}}{1-z_{1}} y_{1},\,\frac{1-t_{1}}{1-z_{1}}x_{0}\right)\,.
\end{equation}

In terms of the ordinary integrals, (\ref{Ge}) has the form
$$
G_{2}^{n}\left(0,\,t_{1} \right)=\frac{1}{\pi^{2}}\left(\frac{\sigma^{2}}{2\pi}\right)^{\frac{n}{2}}\,
\left[t_{1}(1-t_{1})\right]^{-\frac{1}{2}}
$$
$$
\times\int\limits_{0}^{+\infty}\,\int\limits_{0}^{+\infty}d\theta_{1}\, d\theta_{2}\,\exp\left\{-\frac{2}{\sigma^{2}}\left[\frac{\theta_{1}^{2}-\pi^{2}}{t_{1}}+
\frac{\theta_{2}^{2}-\pi^{2}}{(1-t_{1})}\right]\right\}
\,
\sin\left(\frac{4\pi\,\theta_{1}}{\sigma^{2}\,t_{1}}\right)
\sin\left(\frac{4\pi\,\theta_{2}}{\sigma^{2}\,(1-t_{1})}\right)
$$
$$
\times \sinh(\theta_{1})\sinh(\theta_{2})\,\int\limits_{0}^{1}\,dz_{1}\,(1-z_{1})^{\frac{n}{2}}\,
\int\limits_{0}^{+\infty}\int\limits_{0}^{+\infty}\,dx_{0}\,dy_{1}\,\left(x_{0}\,y_{1}\right)^{\frac{n}{4}}
$$
\begin{equation}
   \label{J1}
\times \exp
\left\{-\left(x_{0}+y_{1}+2\left[ (1-z_{1})\cosh \theta_{1} + z_{1}\cosh \theta_{2} \right]\sqrt{x_{0}y_{1}}\, -4z_{1}(1-z_{1})x_{0} \right)\right\}\,.
\end{equation}

After the substitutions
\begin{equation}
   \label{subst}
x_{0}=\rho\cos^{2}\omega\,,\ \ \ y_{1}=\rho\sin^{2}\omega\,,\ \ \ \ \xi=\tan \omega \,,
\end{equation}
 the integrals over $x_{0}$ and $y_{1}$ are transformed into the integral
\begin{equation}
   \label{Intksi}
\int\limits_{0}^{+\infty}\frac{2\,\xi^{1+\frac{n}{2}}}{\left[\xi^{2}+2b\xi+1-4z_{1}(1-z_{1}) \right]^{2+\frac{n}{2}}}\,d\xi\,,
\end{equation}
with
\begin{equation}
   \label{b}
b=(1-z_{1})\cosh \theta_{1}+ z_{1}\cosh \theta_{2}\,.
\end{equation}

Using the table of integrals \cite{(Prud)} (eq. 2.2.9.8), we obtain the result:
$$
G_{2}^{n}\left(0,\,t_{1} \right)=\frac{2}{\pi^{2}}\left(\frac{\sigma^{2}}{2\pi}\right)^{\frac{n}{2}}\,\Gamma^{3}\left(\frac{n+4}{2}\right)\,\Gamma^{-1}\left(n+4\right)\,
\left[t_{1}(1-t_{1})\right]^{-\frac{1}{2}}
$$
$$
\times\int\limits_{0}^{+\infty}\,\int\limits_{0}^{+\infty}d\theta_{1}\, d\theta_{2}\,\exp\left\{-\frac{2}{\sigma^{2}}\left[\frac{\theta_{1}^{2}-\pi^{2}}{t_{1}}+
\frac{\theta_{2}^{2}-\pi^{2}}{(1-t_{1})}\right]\right\}
\,
\sin\left(\frac{4\pi\,\theta_{1}}{\sigma^{2}\,t_{1}}\right)
\sin\left(\frac{4\pi\,\theta_{2}}{\sigma^{2}\,(1-t_{1})}\right)
$$
$$
\times \sinh(\theta_{1})\sinh(\theta_{2})\,\int\limits_{0}^{1}\,dz_{1}\,(1-z_{1})^{\frac{n}{2}}\,
\left[1- 4 z_{1}(1-z_{1})\right]^{-\frac{n+4}{4}}
$$
\begin{equation}
   \label{nodd}
\times \,{ }_{2}F_{1}\left(\frac{n+4}{2},\,\frac{n+4}{2};\,\frac{n+5}{2};\,\frac{1}{2}\left(1-\frac{b}{1- 4 z_{1}(1-z_{1})}\right)\, \right)\,.
\end{equation}

For $n=2k,\ \ \ k=0,\,1\,...\,,$ we can also use  eq. 2.2.9.12 of \cite{(Prud)}, and rewrite the result in the form
$$
G_{2}^{n}\left(0,\,t_{1} \right)=\left(-1\right)^{k+1}\frac{1}{(k+1)!}\frac{1}{2\pi^{2}}\left(\frac{\sigma^{2}}{4\pi}\right)^{k}\Gamma\left(k+2\right)\,
\left[t_{1}(1-t_{1})\right]^{-\frac{1}{2}}
$$
$$
\times\int\limits_{0}^{+\infty}\,\int\limits_{0}^{+\infty}d\theta_{1}\, d\theta_{2}\,\exp\left\{-\frac{2}{\sigma^{2}}\left[\frac{\theta_{1}^{2}-\pi^{2}}{t_{1}}+
\frac{\theta_{2}^{2}-\pi^{2}}{(1-t_{1})}\right]\right\}
\,
\sin\left(\frac{4\pi\,\theta_{1}}{\sigma^{2}\,t_{1}}\right)
\sin\left(\frac{4\pi\,\theta_{2}}{\sigma^{2}\,(1-t_{1})}\right)
$$
$$
\times \sinh(\theta_{1})\sinh(\theta_{2})\,\int\limits_{0}^{1}\,dz_{1}\,(1-z_{1})^{k}\,
$$
\begin{equation}
   \label{even}
\times \,\frac{\partial^{k+1}}{\partial b^{k+1}}\left(\frac{1}{\sqrt{b^{2}-1+ 4 z_{1}(1-z_{1})}}\log\,\frac{b+\sqrt{b^{2}-1+ 4 z_{1}(1-z_{1})}}{b-\sqrt{b^{2}-1+ 4 z_{1}(1-z_{1})}}\right)\,,
\end{equation}
where $b$ is given by (\ref{b}).

In the similar way, we can write the time-ordered (TO) and the out-of-time-ordered (OTO) renormalized four-point correlation functions:
$$
G_{4}^{TO}\left(0,\,t_{1}\,;\ t_{2},\,t_{3} \right)
$$
\begin{equation}
   \label{G4TO}
=\frac{2}{\sigma^{2}}\int\limits_{Diff^{1} ([0,1]) }\delta\left(\frac{\varphi'(1)}{\varphi'(0)}-1 \right)\,\frac{\left(\varphi'(t_{1})\varphi'(0)\right)^{\frac{1}{4}}}{\left |\varphi(t_{1})\right|^{\frac{1}{2}}}\,
\frac{\left(\varphi'(t_{3})\varphi'(t_{2})\right)^{\frac{1}{4}}}{\left |\varphi ( t_{3})-\varphi(t_{2})\right|^{\frac{1}{2}}}\,\exp\left\{\frac{8}{\sigma^{2}}\left(\varphi'(0)\right)\right\}\,\mu_{\sigma}(d\varphi)\,,
\end{equation}
and
$$
G_{4}^{OTO}\left(0,\,t_{2}\,;\ t_{1},\,t_{3} \right)
$$
\begin{equation}
   \label{G4OTO}
=\frac{2}{\sigma^{2}}\int\limits_{Diff^{1} ([0,1]) }\delta\left(\frac{\varphi'(1)}{\varphi'(0)}-1 \right)\,\frac{\left(\varphi'(t_{2})\varphi'(0)\right)^{\frac{1}{4}}}{\left |\varphi(t_{2})\right|^{\frac{1}{2}}}\,
\frac{\left(\varphi'(t_{3})\varphi'(t_{1})\right)^{\frac{1}{4}}}{\left |\varphi ( t_{3})-\varphi(t_{1})\right|^{\frac{1}{2}}}\,\exp\left\{\frac{8}{\sigma^{2}}\left(\varphi'(0)\right)\right\}\,\mu_{\sigma}(d\varphi)\,.
\end{equation}
(In the both equations, we assume $0<t_{1}<t_{2}<t_{3}<1\,.$)

In terms of the functions $\mathcal {E}\,,$ the four-point correlation functions look like
$$
G_{4}^{TO \,(OTO)}\left(0,\,t_{1\,(2)}\,;\ t_{2\,(1)} ,\,t_{3} \right)
=\frac{2}{\sigma^{2}}\,\left[t_{1}(t_{2}-t_{1})(t_{3}-t_{2})(1-t_{3})\right]^{2}\,
$$
$$
\times\,\int\limits_{0}^{1}dz_{1}\int\limits_{z_{1}}^{1}dz_{2}\int\limits_{z_{2}}^{1}dz_{3}\
\chi ^{TO \,(OTO) }\left(z_{1},\,z_{2},\,z_{3}\right)\int\limits_{0}^{+\infty}dx_{0}\int\limits_{0}^{+\infty}
dy_{1}\int\limits_{0}^{+\infty}
dy_{2}\int\limits_{0}^{+\infty}
dy_{3}\,\left[x_{0}\,y_{1}\,y_{2}\,y_{3}\right]^{\frac{5}{4}}
$$
$$
\times\,\exp\left\{\frac{8}{\sigma^{2}}
x_{0} \right\}\,\mathcal {E}_{\sigma\sqrt{t_{1}}}\left(\frac{t_{1}}{z_{1}}\,x_{0},\,\frac{t_{1}}{z_{1}}\,y_{1}\right)\,\mathcal {E}_{\sigma\sqrt{t_{2}-t_{1}}}\left(\frac{t_{2}-t_{1}}{z_{2}-z_{1}}\,y_{1},\,\frac{t_{2}-t_{1}}{z_{2}-z_{1}}\,y_{2}\right)
$$
\begin{equation}
   \label{G4TOOTO}
\times\,
\mathcal {E}_{\sigma\sqrt{t_{3}-t_{2}}}\left(\frac{t_{3}-t_{2}}{z_{3}-z_{2}}\,y_{2},\,\frac{t_{3}-t_{2}}{z_{3}-z_{2}}\,y_{3}\right)\,\mathcal {E}_{\sigma\sqrt{1-t_{3}}}\left(\frac{1-t_{3}}{1-z_{3}}\,y_{3},\,\frac{1-t_{3}}{1-z_{3}}\,x_{0}\right)\,,
\end{equation}
where
\begin{equation}
   \label{chiTO}
\chi ^{TO  }\left(z_{1},\,z_{2},\,z_{3}\right)=
\left[z_{1}\,(z_{2}-z_{1})\,(z_{3}-z_{2})\,(1-z_{3})\right]^{-3}\,
\left[z_{1}\,(z_{3}-z_{2}) \right]^{-\frac{1}{2}}\,,
\end{equation}
and
\begin{equation}
   \label{chiOTO}
\chi ^{OTO }\left(z_{1},\,z_{2},\,z_{3}\right)=
\left[z_{1}\,(z_{2}-z_{1})\,(z_{3}-z_{2})\,(1-z_{3})\right]^{-3}\,
\left[z_{2}\,(z_{3}-z_{1}) \right]^{-\frac{1}{2}}\,.
\end{equation}

The only difference between (\ref{chiTO}) and (\ref{chiOTO}) is in the dependence of the integrands on the variables $z_{i}\,.$

Neither the OTO four-point correlation function nor the TO four-point correlation function has the form of the product of two two-point correlation
functions.

Now we express the four-point correlation functions in the form convenient for numerical analysis.
$$
G_{4}^{TO \,(OTO)}\left(0,\,t_{1\,(2)}\,;\ t_{2\,(1)} ,\,t_{3} \right)
=\left(\frac{2}{\pi^{3}\sigma^{2}}\right)^{2}\,\left[t_{1}(t_{2}-t_{1})(t_{3}-t_{2})(1-t_{3})\right]^{-\frac{1}{2}}
$$
$$
\times\,\int\limits_{0}^{1}dz_{1}\int\limits_{z_{1}}^{1}dz_{2}\int\limits_{z_{2}}^{1}dz_{3}\, \
\tilde{\chi} ^{TO \,(OTO) }\left(z_{1},\,z_{2},\,z_{3}\right)
$$
$$
\times\,\int\limits_{0}^{+\infty}\,d\theta_{1}...\int\limits_{0}^{+\infty}\, \,d\theta_{4}\,\exp\left\{-\frac{2}{\sigma^{2}}\left[\frac{\theta_{1}^{2}-\pi^{2}}{\tau_{1}}+
\frac{\theta_{2}^{2}-\pi^{2}}{(\tau_{2}-\tau_{1})}+
\frac{\theta_{3}^{2}-\pi^{2}}{(\tau_{3}-\tau_{2})}+
\frac{\theta_{4}^{2}-\pi^{2}}{(1-\tau_{3})}\right]\right\}
$$
$$
\times\,
\sin\left(\frac{4\pi\,\theta_{1}}{\sigma^{2}\,\tau_{1}}\right)\,
\sin\left(\frac{4\pi\,\theta_{2}}{\sigma^{2}\,(\tau_{2}-\tau_{1})}\right)\,\sin\left(\frac{4\pi\,\theta_{3}}{\sigma^{2}\,(\tau_{3}-\tau_{2})}\right)\,
$$
$$
\times\,\sinh\theta_{1}\,\sinh\theta_{2}\,\sinh\theta_{3}\,\sinh\theta_{4}
\int\limits_{0}^{+\infty}dx_{0}\int\limits_{0}^{+\infty}
dy_{1}\int\limits_{0}^{+\infty}
dy_{2}\int\limits_{0}^{+\infty}
dy_{3}\,\left[x_{0}\,y_{1}\,y_{2}\,y_{3}\right]^{\frac{1}{4}}
$$
$$
\times \exp\left\{4x_{0}-\frac{1}{z_{1}}\left[x_{0}+y_{1}+2\sqrt{x_{0}y_{1}}\cosh \theta_{1} \right]
-\frac{1}{z_{2}-z_{1}}\left[y_{1}+y_{2}+2\sqrt{y_{1}y_{2}}\cosh \theta_{2} \right]\right\}
$$
\begin{equation}
\label{4int}
\times \exp\left\{-\frac{1}{z_{3}-z_{2}}\left[y_{2}+y_{3}+2\sqrt{y_{2}y_{3}}\cosh \theta_{3} \right]
-\frac{1}{1-z_{3}}\left[y_{3}+x_{0}+2\sqrt{y_{3}x_{0}}\cosh \theta_{4} \right] \right\}\,,
\end{equation}
where
$$
\tilde{\chi} ^{TO  }\left(z_{1},\,z_{2},\,z_{3}\right)=
\left[z_{1}\,(z_{2}-z_{1})\,(z_{3}-z_{2})\,(1-z_{3})\right]^{-2}\,
\left[z_{1}\,(z_{3}-z_{2}) \right]^{-\frac{1}{2}}\,,
$$
and
$$
\tilde{\chi} ^{OTO }\left(z_{1},\,z_{2},\,z_{3}\right)=
\left[z_{1}\,(z_{2}-z_{1})\,(z_{3}-z_{2})\,(1-z_{3})\right]^{-2}\,
\left[z_{2}\,(z_{3}-z_{1}) \right]^{-\frac{1}{2}}\,.
$$

Using the general rule (\ref{RuleK}) we can write any $n-$point correlator in the same way.

 It would be interesting to study the possible chaotic behaviour of the model with correlation functions (\ref{G4TOOTO}).

\section{Concluding remarks}
\label{sec:concl}

The paper completes the elaboration of  Schwarzian functional integrals calculus initiated  in  \cite{(BShExact)} and  \cite{(BShCorrel)}.
Our approach is  mathematically rigorous and does not contain any unproved conjectures.
It does not use perturbation theory and does not appeal for the experience from other physical models.
The general rules derived here give a straightforward scheme of calculation of functional integrals that is applicable in a wide class of problems.

The great merit of our approach is that it reduces  various problems to the evaluation of the functional integral (\ref{E})
only. It is evaluated explicitly and is written in the form of the ordinary integral (\ref{Eint}).
Hence, the evaluation of other functional integrals of the form (\ref{RuleK}) leads also to ordinary (multiple) integrals.

The results of functional integration over the groups of diffeomorphisms of a circle (even of infinite radius) and of an infinite straight line differ significantly, contradicting a naive physical intuition.
Indeed, using the method described in this paper, one can easily evaluate functional integrals over the group   $ Diff^{1}_{+}([-T,\,+T])\,.$ Then there are two options (corresponding to two different theories) to pass to the limit
$T\rightarrow \infty\,:$ keeping values of $\varphi'(-T)$ and $\varphi'(+T)$ arbitrary and independent of each other; and with the ends of the interval being glued $\varphi'(-T)=\varphi'(+T)\,.$ In the first case, one obtains the results of section \ref{sec:diffR}, while the second option leads to the results of section \ref{sec:correl}.
It is quite natural that functional integrations over different integration spaces ($ Diff^{1}_{+}(S^{1})$ and $Diff^{1}_{+}(\textbf{R})$)  give different results for the same integrand.
(\emph{Ouroboros} is not the same as an ordinary snake!)

Gluing the ends of the interval together turns the measure $\mu_{\sigma}(d\varphi)$ and
the corresponding Wiener measure into the conditional  measures.
The fact that Markov property is violated for the measure (\ref{MeasureS}) on the group of diffeomorphisms of the circle has very important consequences. In particular,
the regions with $t>t_{1}$ (for $G_{2}\left(0,\,t_{1} \right)$), and with $t>t_{3}$ (for $
G_{4}\left(0,\,t_{1\,(2)}\,;\ t_{2\,(1)} ,\,t_{3} \right)$ ) give the nonzero inputs into the integrals for correlation functions\footnote{ Considering  $t $ as a time variable, one could say that "the present" is influenced by "the future". }. And also, neither time-ordered nor out-of-time-ordered four-point  correlation function is represented in the form of a product of two two-point correlation functions.
Actually, these properties , being in contrast to those of (\ref{Bksi}), (\ref{gTO}), (\ref{gOTO}), distinguish the results of integration over the group of diffeomorphisms of the circle and those over the groups of
diffeomorphisms of the interval or of the real axis.

In section \ref{sec:diffR}, we demonstrate that our results for two-point and four-point correlation functions coincide with those in  \cite{(BAK1)}.
The authors of \cite{(MTV)} stated the coincidence of their results with those of \cite{(BAK1)}, \cite{(BAK2)} at p.19, after eq. (4.11), and at p.25, footnote 14.  The authors of
\cite{(KitSuh2)} also claimed (p.33, end of section 5) that in the so-called Schwarzian limit they obtained "an expression for the Euclidean correlator that coincides with (4.10) in \cite{(MTV)} up to a constant factor and is also consistent with equations (22), (23) in \cite{(BAK1)}".

Thus we can conclude that the results (\ref{gnResult}) and (\ref{gTO}), (\ref{gOTO}) of
functional integration over the group $Diff^{1}_{+}(\textbf{R})$ are equivalent to the results for two-point correlation function and four-point correlation functions in \cite{(BAK1)} and in \cite{(MTV)}, and also for two-point function in \cite{(KitSuh2)}.

In
\cite{(Yang)}, correlation functions are represented as integrals over the coordinates of a circle on the hyperbolic plane $H_{2}$ with closed chains of propagators in integrands\footnote{See equation (6.3), (6.5) in \cite{(Yang)}.}. Thus correlation functions have qualitatively similar (non-Markov) behaviour
as that obtained by the integration over the group of diffeomorphisms of the circle.

However, before the identification of the inputs of the problems a detailed comparison of the results for correlation functions has no sense.
The point is that
functional integrals in \cite{(Yang)} are written as the sums over the eigenstates of the Hamiltonian of the certain physical model. At the special limiting values of parameters, the action of the model is reduced to the Schwarzian action, and the sum over the eigenstates for the partition function leads to the Schwarzian spectral density $\varrho(E)=\sinh (2\pi\sqrt{E})$ in this limit.

Although, the circumstantial way of calculation in \cite{(Yang)} seems to be equivalent to Schwarzian functional integration "at a physical level of rigor", a mathematically rigorous proof is an open problem. It concerns the explicit form of the measure and the specification of the integration space in the limiting case (see the discussion in \cite{(Yang)}), and also matching the $SL(2,\textbf{R})$ gauge-fixing procedures.

In particular,  the measure of functional integrals in the physical model used in \cite{(Yang)} as the basis for calculations is
$$
\exp\left\{-\frac{1}{2}\int\limits_{0}^{\tau}\left(\frac{\dot{x}^{2}+\dot{y}^{2}}{y^{2}}+q\,\frac{\dot{x}}{y} \right)\,dt\right\}\,dx\,dy
$$
It can be represented as the measure with depending on each other variables $\eta$ and $\varphi$
\begin{equation}
\label{nondprod}
w_{\frac{1}{q}}\left(d\eta(x\,,\,y)\right)\ \mu \left(d\varphi(x\,,\,y)\right) \,.
\end{equation}
If (\ref{nondprod}) were the direct product of the measures with independent variables, its limit at $q\rightarrow\infty$ would have been the measure
(\ref{Measure}). However, the real problem is more difficult.

The note concerns not only \cite{(Yang)} but all other approaches where circumstantial ways of calculation are used.

On the other hand, in this paper, we integrate over the definite functional spaces using rigorously proved properties of the measures on these spaces. Therefore we believe
that the results obtained by our method can be considered as
a frame of reference for studying path integrals in various physical models
including those in the finite-cutoff JT gravity
\cite{(IKTVfincut)},  \cite{(SYfincut)}.

\end{document}